\documentclass{ws-mpla}

\usepackage[super]{cite}
\usepackage{xcolor}
\usepackage[verbose,hypertexnames=false]{hyperref}
\hypersetup{colorlinks=false,allbordercolors=blue,pdfborderstyle={/S/U/W 1}}

\newcommand{\gdualn}[1]{\overset{\:{}^{{}^{\boldsymbol{\neg}}}}{\smash[t]{#1}}} 
\newcommand{\elko}{\xi}

\begin{document}
	
	
	\catchline{}{}{}{}{}
	
	\title{On the structure of interactions of mass dimension one fermions:\\ a functional renormalization group perspective}

	\author{Gustavo P. de Brito}
	
	\address{Universidade Estadual Paulista (Unesp), Av.~Dr.~Ariberto Pereira da Cunha, 333\\ Guaratinguet\'a, S\~ao Paulo, 12516-410, Brazil.
		\\
		gp.brito@unesp.br}

	\maketitle
	
	
	\begin{abstract}
		In this paper, we provide the first systematic investigation of renormalization group properties of mass dimension one fermions described by ELKO spinors. By construction, ELKOs must be neutral under any Standard Model charge, therefore, providing a natural candidate for dark matter. We consider two versions of scalar-ELKO systems: the first characterized by a derivative Yukawa-like interaction, while the second involves ELKO four-fermion interactions as well as a scalar-ELKO portal. We also considered a system composed of ELKOs interacting with an Abelian gauge field via Pauli-like term. In all cases, we identified the minimal set of interactions that are required by a consistent renormalization group flow, and we discussed the possibility of constructing UV-complete trajectories based on asymptotic freedom. We used the functional renormalization group as a method of investigation.
	\end{abstract}
	
	\keywords{ELKO; Renormalization Group;}
	
	
	\vspace*{.5cm}
	\noindent \textit{This paper is dedicated to Dharam Vir Ahluwalia, for his courage in pursuing ideas beyond the existing paradigms and expanding the horizons of fermionic fields.}
	
	\section{Introduction}	
	
	Since the foundational works by Weinberg and other pioneers of quantum field theory on the irreducible representations of the Poincaré group, see for instance ~\cite{Weinberg:1964cn,Weinberg:1995mt}~, the prevailing point of view is that relativistic massive spin-1/2 particles must be described either by Dirac or Majorana spinors. Around two decades ago, this point of view was challenged in two seminal papers by Ahluwalia and Grumiller ~\cite{Ahluwalia:2004ab,Ahluwalia:2004sz}~, where they reported the discovery of a new type of spinor belonging to the $(1/2,0) \oplus (0,1/2)$ representation of the Lorentz group (same as Dirac and Majorana spinors), but with completely distinct dynamical properties. This new type of spinor was introduced as eigenstates of the charge conjugation operator with eigenvalues $\pm 1$. The new spinor was called ELKO as an acronym for \textit{\textbf{E}igenspinoren des \textbf{L}adungs\textbf{k}onjugations\textbf{o}perators}, which is the German translation of \textit{Eigenspinors of Charge Conjugation Operator}.
	
	One of the main features of the ELKO is that its dynamics is not described by the Dirac equations ~\cite{Ahluwalia:2004ab,Ahluwalia:2004sz}~. Instead, its dynamical properties (at least as a free field) are entirely encoded in the Klein-Gordon equation. Despite this feature, the ELKO is still consistent with fermionic statistics. However, thanks to its ``Klein-Gordon dynamics'', quantum fields constructed with ELKO spinors as coefficients of creation and annihilation operators have canonical mass dimension one, not three-halves as the other spinors that appear in particle physics. Therefore, the ELKO paradigm describes \textit{mass dimension one fermions}.
	
	In the first two decades after the discovery of the ELKO, there has been a significant effort on understanding the mathematical and foundational properties of this new type of spinor, ranging from its classification according with Lounesto classes of spinors ~\cite{daRocha:2005ti,daRocha:2008we,HoffdaSilva:2012uke,HoffdaSilva:2017waf} to its consolidation as an irreducible representation of the Poincaré group ~\cite{Ahluwalia:2022yvk,Ahluwalia:2023slc}~. See also ~\cite{Ahluwalia:2022ttu} for an almost updated review on the achievements of the ELKO paradigm.
	In particular, it took several years until the understanding that the ELKO fits an irreducible representation featuring a two-fold degeneracy already envisioned by Wigner in the early days of quantum field theory ~\cite{Ahluwalia:2023slc}~. This was a crucial milestone in the theory of ELKOs, finally allowing the definition of a dual-spinor structure that is compatible with a Lorentz covariant completeness relation ~\cite{Ahluwalia:2023slc}~, without appealing the so-called $\tau$-deformation introduced in ~\cite{Ahluwalia:2016rwl}~.
	
	On physical grounds, the ELKO is a natural candidate to describe dark matter ~\cite{Ahluwalia:2004ab,Ahluwalia:2004sz,Agarwal:2014oaa,Pereira:2018xyl,Pereira:2021dkn,Moura:2021rmf}~. Due to its definition as an eigenstate of the charge conjugation operator, the ELKO is forced to be neutral under the Standard Model gauge group. In particular, it cannot interact electromagnetically via minimal couplings. Therefore, if we restrict ourselves to perturbatively renormalizable interactions, the ELKO can only communicate with Standard Model fields via Higgs-portal interaction or via a Pauli-like interaction involving the hyper-charge gauge field ~\cite{Ahluwalia:2004ab,Ahluwalia:2004sz}~. Of course, the ELKO also interacts with gravity ~\cite{Boehmer:2006qq,BuenoRogerio:2019zvz}~. Cosmological implications of ELKOs were discussed in ~\cite{Boehmer:2007dh,Boehmer:2008ah,Boehmer:2008rz,Boehmer:2009aw,Boehmer:2010ma,HoffdaSilva:2014tth,Pereira:2014wta,S:2014woy,Pereira:2014pqa,Pereira:2016emd,Pereira:2016eez,Pereira:2017bvq,Pereira:2017efk,Pereira:2018hir,Pereira:2020ogo}~. For particle phenomenology implications, see ~\cite{Dias:2010aa,Alves:2014kta,Alves:2017joy,Duarte:2020svn}~.
	
	In the recent years, there has been an increasing interest in understanding the structure of interactions involving ELKOs, as well as their quantum properties. The first studies of quantum corrections of correlation functions involving ELKOs were done in ~\cite{deBrito:2019hih,Nayak:2020ntm}~, where the first focuses on one-loop renormalizability of ELKO self-interactions, while the second explores the impact of ELKO loops on the value of the Higgs mass. Investigation of radiative corrections and the renormalizability of ELKOs minimally coupled (via Pauli-like interaction) with abelian gauge fields was done in ~\cite{Carvalho:2023btn}~.
	
	The type of interactions considered in ~\cite{deBrito:2019hih,Nayak:2020ntm,Carvalho:2023btn} share a common feature: they are non-Hermitian interactions. Instead, as discussed in ~\cite{Ahluwalia:2023slc}~, they can be formulated as pseudo-Hermitian interactions. More recently, the authors of ~\cite{deGracia:2024umr,deGracia:2025omq} introduced a derivative Yukawa-type interaction involving the ELKO and the Higgs boson, which is compatible with hermiticity. However, since this interaction cannot be extended to include a Higgs doublet, it is unclear how to make this proposal compatible with the symmetric phase of the Standard Model.
	
	In this paper, we will perform a systematic investigation of the structure of interactions involving ELKOs under a Wilsonian renormalization group perspective. As a method of investigation, we employ the functional renormalization group (FRG) toolbox to explore what kind of interaction are necessarily generated once we integrate out high-momentum modes ~\cite{Dupuis:2020fhh}~. The analysis performed in this paper includes self-interactions, interactions with a scalar, and an Abelian gauge fields.
	
	This paper is organized as follows: In Sec. \ref{sec:MethodSetup}, we provide a brief introduction to the FRG and we define the setup used in this paper; In Sec. \ref{sec:ELKO+Scalar}, we discuss the renormalization group flow of two versions of scalar-ELKO systems; In Sec. \ref{sec:ELKO+Gauge}, we discuss the renormalization group flow of a gauge-ELKO system with Pauli-like interaction; In Sec. \ref{sec:Conclusion}, we present our final remarks. We discuss some technical details about the behavior of renormalization group trajectories around a free fixed point in the \ref{app:LO-Flow}. 
	
	\section{Methodology and setup \label{sec:MethodSetup}}
	
	\subsection{Brief introduction to the functional renormalization group}	
	
	Since this is the first paper using the functional renormalization group to study ELKOs, it is convenient to have a brief introduction about this tool to make this paper accessible to readers that are not familiar with it. For the part of the audience that is familiar with the functional renormalization group, but not with ELKOs, Sec. 2 of Ref. ~\cite{Ahluwalia:2023slc} offers an up-to-date discussion on this topic.
	
	The functional renormalization group (FRG) is a practical implementation of the Wilsonian ideas on renormalization group, see ~\cite{Dupuis:2020fhh} for the most recent review on the topic. The main idea of the Wilsonian renormalization is the implementation of a coarse-graining procedure, at the level of functional integrals in Euclidean quantum field theories, by integrating high-momentum modes while keeping low-momentum modes untouched. This allows us to define an effective low-momentum dynamics with the details of high-momentum modes absorbed into a redefinition of couplings that characterize the model. 
	
	Within the FRG, the implementation of this coarse-graining procedure is done by introducing an infrared (IR) cutoff scale $k$ that separates low-momentum from high-momentum modes. At the practical level, this cutoff is introduced by replacing the bare action according to
	\begin{equation}
		S[\phi] \mapsto S_k[\phi] = S[\phi] + \frac{1}{2} \int \!\! d^4x \, \phi(x) \,R_k(\Delta) \,\phi(x) \,,
	\end{equation}
	where $R_k(\Delta)$ is a regulator function, $\phi(x)$ is a generic field, and $\Delta = - \partial_\mu \partial^\mu$ is (minus) the Laplace operator whose eigenvalues ($q^2 = q_\mu q^\mu$, where $q^\mu$ is an Euclidean 4-momentum vector) are used to define what is a low-momentum or high-momentum mode in comparison with $k^2$. If the field $\phi(x)$ carries one or more indices, then the regulator function $R_k(\Delta)$ should also carry a tensor structure accordingly. 
	
	In general, the regulator function $R_k$ is defined by the following requirements: 
	\begin{enumerate}
		\item[i)] for a fixed value of $k$, $R_k(z)$ should decrease with $z$; 
		\item[ii)] for a fixed value of $z$, $R_k(z)$ should increase with $k$; 
		\item[iii)] $\lim_{k\to0} R_k(z) = 0$ for any value of $z$; 
		\item[iv)] For $z>k^2$, $R_k(z)$ should go to zero sufficiently fast when increasing $z$. 
	\end{enumerate}
	Typically, we also demand a normalization of the form $R_k(0) = Z_k k^2$, where $Z_k$ is a wave function renormalization factor. This set of requirements is sufficient to ensure the suppression of low-momentum modes while keeping high-momentum modes unsuppressed in a functional integral regularized with an IR cutoff $k$. A common choice for $R_k$ is the Litim regulator ~\cite{Litim:2000ci,Litim:2001up}~, defined as follows 
	\begin{equation}
		R_k(z) = Z_k (k^2 - z) \theta(1 - z/k^2) \,,
	\end{equation}
	where $\theta$ is the Heaviside function. This is a convenient choice of regulator since it allows analytical evaluation of certain integrals. 
	
	The central object of the FRG is the flowing action $\Gamma_k[\phi]$, which is defined to capture the effective dynamics of the system once we integrate modes with momentum above $k$, see for instance ~\cite{Dupuis:2020fhh}~ for a more technical definition of $\Gamma_k[\phi]$. Moreover, $\Gamma_k[\phi]$ is defined such that it interpolates between the bare action $S[\phi]$ at $k\to \Lambda_\text{UV}$ (with $\Lambda_\text{UV}$ being an ultraviolet (UV) cutoff) and the full effective action $\Gamma[\phi]$ at $k \to 0$. 
	
	The flowing action satisfy an exact renormalization group equation, known as the Wetterich equation, which is given by ~\cite{Wetterich:1992yh,Ellwanger:1993mw,Morris:1993qb}
	\begin{equation}\label{eq:WetterichEq}
		k \partial_k \Gamma_k[\phi] = \frac{1}{2} \text{STr} \left[ \big(\Gamma_k^{(2)} + R_k\big)^{-1} k \partial_k R_k \right] \,,
	\end{equation}
	where $\text{STr}$ is the super-trace, which traces over internal and spacetime indices, with an additional negative sign for fermionic fields, and $\Gamma_k^{(2)}$ is a scale-dependent 2-point function defined as the second functional derivative of $\Gamma_k[\phi]$ with respect to the field $\phi$. 
	
	In general, we do not know how to find exact solutions to the Wetterich equation. However, by applying approximation strategies, based on truncations of $\Gamma_k[\phi]$, we can still extract relevant information from the Wetterich equation. The general strategy starts with an \textit{ansatz} for flowing action, which is defined by expanding $\Gamma_k[\phi]$ in a set of linearly independent operators compatible with the symmetries of the system under consideration. 
	The schematic form of a truncation for the flowing action is given by
	\begin{equation}\label{eq:truncation_general}
		\Gamma_k[\phi] = \sum_i k^{-\Delta_i} g_i(k) \mathcal{O}_i[\phi] \,,
	\end{equation}
	where the $\mathcal{O}_i[\phi]$'s are operators compatible with the symmetries of the problem, $g_i(k)$'s are scale-dependent (dimensionless) couplings, and $\Delta_i$ denotes the canonical mass dimension of the corresponding operator $\mathcal{O}_i[\phi]$.
	Within a given truncation, the problem of solving the scale-dependence of $\Gamma_k[\phi]$ is converted into solving the scale-dependence of the set of couplings $\{g_i(k)\}$, which requires the evaluation of the corresponding beta functions $\beta_{g_i} = k\partial_k g_i(k)$. By factoring out $k^{-\Delta_i}$ in each term of the truncation, one can ensure that the scale-dependence of the corresponding couplings are described by an autonomous dynamical system.
	
	The typical definition of a truncation for $\Gamma_k[\phi]$ allows us to identify one or more beta functions from the flow of vertices projected into a given momentum configuration. The basic strategy goes as follows: Let $\Gamma^{(n)}_{k}$ be a (scale-dependent) vertex defined by taking $n$-functional derivatives of $\Gamma_k[\phi]$ and then taking $\phi=0$. From \eqref{eq:truncation_general}, the vertex $\Gamma^{(n)}_k$ takes the form
	\begin{equation}\label{eq:vertex_general}
		\Gamma^{(n)}_k = \sum_{i} k^{-\Delta_i} g_i(k) \, \mathcal{O}_i^{(n)}\,.
	\end{equation}
	where $\mathcal{O}_i^{(n)}$ is defined by taking $n$-functional derivatives of $\mathcal{O}_i[\phi]$, and then setting $\phi=0$. Typically, $\Gamma^{(n)}_k$ involves a subset of couplings defined in the original truncation, as only those terms satisfying $\mathcal{O}_i^{(n)} \neq 0$ survives in \eqref{eq:vertex_general}. To select a single coupling, it is useful to define a projector such that $\mathcal{P}_i \circ \mathcal{O}_j^{(n)} = \delta_{ij}$, thus leading to
	\begin{equation}
		k^{-\Delta_i} g_i(k) =  \mathcal{P}_i \circ\Gamma^{(n)}_k \,.
	\end{equation}
	In general, this projection operation may involve tensor contractions as well as the choice of a particular momentum configuration. Therefore, to evaluate the beta function of a given coupling $g_i(k)$, associated with a vertex $\Gamma_k^{(n)}$, one can take $n$-functional derivatives of the Wetterich equation and then act on both sides with the corresponding projector $\mathcal{P}_i$. Schematically, one has
	\begin{equation}
		\beta_{g_i} = \Delta_i \,g_i + k^{\Delta_i} \mathcal{P}_i \circ \bigg[ \frac{\delta^n}{\delta \phi^n} 
		\left( \frac{1}{2} \text{STr} \left[ \big(\Gamma_k^{(2)} + R_k\big)^{-1} k \partial_k R_k \right)
		\right)\bigg]_{\phi = 0} \,,
	\end{equation}
	where the first term drives the scale-dependence due to the canonical mass dimension. Throughout this paper, we will work with the so-called one-loop approximation of the FRG, where we neglect the scale-dependence of the wave-function renormalization factors coming from $k \partial_k R_k$.
	
	Since the truncation strategy typically does not involve a polynomial expansion on the interaction couplings, it allow us to extract non-perturbative information about the renormalization group flow of strongly correlated systems. This strategy has been successfully employed in several branches of theoretical physics, ranging from condensed matter to quantum gravity ~\cite{Dupuis:2020fhh}~.
	
	\subsection{Defining the setup}	
	
	The general setup used in this paper is defined by a truncation for $\Gamma_k$ involving ELKO spinors potentially coupled with a real scalar field and an Abelian gauge field. 
	
	The reason why we focus on ELKO interactions with a scalar and an Abelian gauge field is that these are the possible perturbatively renormalizable portals between the Standard Model of particle physics and a possible dark sector described by ELKOs. Since, by construction, the ELKO has to be a singlet under the Standard Model gauge group, we cannot construct any perturbatively renormalizable interactions between ELKOs and the other Standard Model fields. For simplicity, we will consider a single real scalar field instead of the Higgs doublet. This simplification should not affect most of the qualitative results presented in this paper. The exception is the derivative Yukawa-like interaction, which cannot be readily extended to include a Higgs doublet.
	
	Being more precise about the setup, we will consider a system with $N_\text{E}$ ELKO spinors denoted as $\elko_a(x)$ (where $a$ is a ``flavor'' index ranging from 1 to $N_\text{E}$), a single real scalar field denoted as $\varphi(x)$, and a single Abelian gauge field denoted as $A_\mu(x)$. The dual ELKO spinor is denoted as $\gdualn{\elko}_a(x)$. For a free theory, the dual ELKO is defined as $\gdualn{\elko}_a(x) = [i\, m^{-1}  \gamma^\mu\partial_\mu\elko_a(x)]^\dagger \gamma^0$, where $m$ is a fiducial mass scale, usually identified with the ELKO mass parameter. For the interacting system, $\gdualn{\elko}_a(x)$ should be treated as an independent field.
	
	Throughout the next sections, we are going to work with the following truncation for the flowing action
	\begin{eqnarray}
		\Gamma_k[\elko,\gdualn{\elko},\varphi,A] = \Gamma_k^{\text{ELKO}}[\elko,\gdualn{\elko}] +  \Gamma_k^{\text{scalar}}[\varphi] +  \Gamma_k^{\text{gauge}}[A] + \Gamma_k^{\text{int.}}[\elko,\gdualn{\elko},\varphi,A] \,,
	\end{eqnarray}
	where
	\begin{subequations}
		\begin{equation}
			\Gamma_k^{\text{ELKO}}[\elko,\gdualn{\elko}] = \int \! d^4x \, \left( Z_{\elko}(k) \,\partial_\mu \gdualn{\elko}_a \partial^\mu \elko_a +  Z_{\elko}(k)\, \bar{m}_{\elko}^2(k)   \gdualn{\elko}_a \elko_a  \right) \,,
		\end{equation}
		\begin{equation}
			\Gamma_k^{\text{scalar}}[\varphi] = \int \! d^4x \, \bigg( \frac{1}{2} Z_{\varphi}(k)\, \partial_\mu \varphi \,\partial^\mu \varphi + \frac{1}{2} Z_{\varphi}(k) \, \bar{m}_{\varphi}^2(k) \varphi^2  \bigg) \,,
		\end{equation}
		\begin{equation}
			\Gamma_k^{\text{gauge}}[A] = \int \! d^4x  \bigg( \frac{1}{4} Z_{A}(k) \, F_{\mu\nu} F^{\mu\nu} + \frac{1}{2\alpha} Z_{A}(k) \, (\partial_\mu A^\mu)^2  \bigg)  \,,
		\end{equation}
	\end{subequations}
	and $\Gamma_k^{\text{int.}}[\elko,\gdualn{\elko},\varphi,A]$ contains the interaction terms that will explicitly defined in the next sections. The parameters $\bar{m}_{\elko}^2(k)$ and $\bar{m}_{\varphi}^2(k)$ are scale-dependent mass parameters, while $Z_{\elko}(k)$, $Z_{\varphi}(k)$ and $Z_{A}(k)$ denote wave-function renormalization factors. 
	The flow of the wave-function renormalization factors is encoded on the corresponding anomalous dimensions $\eta_\elko = -Z_\elko(k)^{-1} \, k\partial_k Z_\elko(k)$, $\eta_\varphi = -Z_\varphi(k)^{-1} \, k\partial_k Z_\varphi(k)$ and $\eta_A = -Z_A(k)^{-1} \, k\partial_k Z_A(k)$.
	We define dimensionless versions of the the mass parameters according to $m_{\varphi}^2(k) = k^{-2} \, \bar{m}_{\varphi}(k)^2$ and $m_{\elko}^2(k) = k^{-2} \,  \bar{m}_{\elko}^2(k)$.
	Finally, $\alpha$ is a gauge-fixing parameter. For convenience, we will focus on the Landau gauge choice $\alpha \to 0$, which selects the transverse part of the gauge field propagator.
	
	It is important to observe that the truncation of $\Gamma_k$ was defined with Euclidean signature, in accordance with the standard formulation of the FRG \footnote{The standard formulation of the FRG relies on spaces with Euclidean signature. This requirement comes from the fact that the implementation of the Wilsonian coarse-graining rely on intrinsically defined momentum shells as a way to order the modes. In the last few years, there was a significant progress on the formulation of a Lorentzian version of the FRG ~\cite{Fehre:2021eob,Braun:2022mgx,DAngelo:2022vsh,Banerjee:2022xvi,DAngelo:2023wje,DAngelo:2023tis,Banerjee:2024tap,Pawlowski:2025etp,Kher:2025rve}~, especially in the context of quantum gravity.}.
	
	To avoid cumbersome notation, throughout the upcoming sections we will represent the couplings without their explicit dependence on $k$. However, it should be implicitly understood that we are referring to scale-dependent couplings.
	Moreover, we will often represent scale-dependent $n$-point vertices as $\Gamma_{\phi_1\cdots \phi_n}^{(n)}$, where the subscript indicates the fields that we are taking functional derivatives.
	
	In this paper, we will define $\Gamma_k^{\text{int.}}[\elko,\gdualn{\elko},\varphi,A]$ such that it includes only perturbatively renormalizable interactions. Higher-order interactions are always generated by the Wilsonian flow. However, since we are interested in discussing perturbative aspects of the flow, we can safely neglect such higher-order terms since the effects of the corresponding couplings are generically suppressed by their negative mass dimension, as long we stay within a perturbative regime.
	
	All the results presented in this paper were derived by a self-written \textit{Mathematica} code ~\cite{Mathematica} based on the packages \textit{xAct} ~\cite{Martin-Garcia:2007bqa,Nutma:2013zea,Martin-Garcia:2008ysv}~, \textit{DoFun} ~\cite{Huber:2011qr,Huber:2019dkb}~, and \textit{FormTracer} ~\cite{Cyrol:2016zqb}~. Large tensor contractions were done with \textit{FORM} ~\cite{Vermaseren:2000nd,Kuipers:2012rf}~. Due to the fully automated nature of these calculations, we will only report the main expressions.
	
	\section{ELKO interacting with scalar fields \label{sec:ELKO+Scalar}}
	
	In this section, we investigate the structure of interactions involving ELKOs and a scalar field. In this case, the Abelian gauge field completely decouples from the rest of the system, and one can just neglect $\Gamma_k^\text{gauge}$ in our truncation. 
	
	\subsection{Derivative Yukawa-like interaction \label{subsec:ScalarELKO_1}}
	
	In the first part of this section, we will discuss the derivative Yukawa-like interaction introduced in ~\cite{deGracia:2024umr}~. We will also consider a quartic self-interaction term in the scalar sector, as this will be unavoidably generated by the renormalization group flow. 
	
	In terms of a truncation for $\Gamma_k$, this setup corresponds to the following interaction term
	\begin{equation}
		\Gamma_k^{\text{int.}}[\elko,\gdualn{\elko},\varphi] =
		\int d^4 x \left(  Z_\varphi^{1/2} Z_\elko \,i\,g_{\varphi\elko} \, \varphi \gdualn{\elko}_a \gamma^\mu\partial_\mu \elko_a + \frac{1}{4!} Z_\varphi^2 \lambda_{\varphi} \varphi^4 \right) \,,
	\end{equation}
	where $g_{\varphi\elko}$ and $\lambda_{\varphi}$ denote scale-dependent couplings. 
	
	The first term in $\Gamma_k^{\text{int.}}[\elko,\gdualn{\elko},\varphi]$ is the derivative Yukawa-like interaction introduced in ~\cite{deGracia:2024umr}~, being motivated by the fact that this is the only interaction involving ELKOs simultaneously compatible with hermiticity and perturbative renormalizability. However, this interaction cannot be extended to accommodate a Higgs doublet; thus, it is unclear how to make it compatible with the symmetric phase of the Standard Model.
	
	Now, let us turn to the renormalization group flow of $g_{\varphi\elko}$ and $\lambda_{\varphi}$, which are associated to the flow of $\Gamma_{\varphi\elko\gdualn{\elko}}^{(3)}$ and $\Gamma_{\varphi\varphi\varphi\varphi}^{(4)}$. 
	Schematically, we can write
	\begin{subequations}
		\begin{equation}
			\beta_{g_{\varphi\elko}} = \left( \frac{1}{2}\eta_\varphi + \eta_\elko \right) g_{\varphi\elko} + \mathcal{P}_{g_{\varphi\elko}} \circ \Big( k\partial_k \Gamma_{\varphi\elko\gdualn{\elko}}^{(3)} \Big) \,,
		\end{equation}
		\begin{equation}
			\beta_{\lambda_{\varphi}} = 2 \,\eta_\varphi  \lambda_{\varphi} + \mathcal{P}_{\lambda_{\varphi}} \circ \Big( k\partial_k \Gamma_{\varphi\varphi\varphi\varphi}^{(4)} \Big) \,.
		\end{equation}
	\end{subequations}
	We use the flow of the 2-point functions $\Gamma_{\varphi\varphi}^{(2)}$ and $\Gamma_{\elko\gdualn{\elko}}^{(2)}$ to derive the anomalous dimensions and the beta functions for the (dimensionless) mass parameters $m_\phi^2 $ and $m_\elko^2$. The explicit results are given by
	\begin{subequations}
		\begin{equation}
			\eta_\varphi = - \frac{N_\text{E}\, g_{\varphi\elko}^2}{4\pi^2 (1+m_\elko^2)^4} \,,
		\end{equation}
		\begin{equation}
			\eta_\elko = - \frac{g_{\varphi\elko}^2}{\pi^2 (1+m_\elko^2)(1+m_\varphi^2)^2} \,,
		\end{equation}
	\end{subequations}
	for the anomalous dimensions, and
	\begin{subequations}
		\begin{equation}
			\beta_{m_\varphi^2} = -2 \,m_\varphi^2 - \frac{\lambda_{\varphi}}{32\pi^2 (1+m_\varphi^2)^2} - \frac{N_\text{E}\, (4+4\,m_\elko^2+3\,m_\varphi^2) \,g_{\varphi\elko}^2}{12\pi^2(1+m_\elko^2)^4} \,,
		\end{equation}
		\begin{equation}
			\beta_{m_\elko^2} = -2\,m_\elko^2 - \frac{m_\elko^2 \,g_\varphi^2}{\pi^2 (1+m_\elko^2)(1+m_\varphi^2)^2} \,,
		\end{equation}
	\end{subequations}
	for the mass parameters.
	Going back to the beta functions of $g_{\varphi\elko}$ and $\lambda_\varphi$, we find
	\begin{subequations}
		\begin{equation}
			\! \beta_{g_{\varphi\elko}} = - \left( \frac{24}{(1+m_\elko^2)(1+m_\varphi^2)^2} +
			\frac{3+m_\elko^2+2\,m_\varphi^2}{(1+m_\elko^2)^3(1+m_\varphi^2)^2} +
			\frac{3N_\text{E}}{(1+m_\elko^2)^4}
			\right) \frac{g_{\varphi\elko}^3}{24\pi^2} \,,
		\end{equation}
		\begin{equation}
			\beta_{\lambda_\varphi} =  \frac{3 \lambda_\varphi^2}{16 \pi^2 (1+m_\varphi^2)^3} 
			- \frac{N_\text{E}\, g_{\varphi\elko}^2 \lambda_\varphi}{2\pi^2(1+m_\elko^2)^4}
			-  \frac{3 N_\text{E} \, g_{\varphi\elko}^4}{\pi^2 (1+m_\elko^2)^5} \,.
		\end{equation}
	\end{subequations}
	
	In addition to this set of renormalization group equations, we also looked at the flow of the 4-point function $\Gamma_{\elko\elko\gdualn{\elko}\gdualn{\elko}}^{(4)}$ and $\Gamma_{\varphi\varphi\elko\gdualn{\elko}}^{(4)}$ to explicitly verify whether operators of the type $\varphi^2 \gdualn{\elko}\elko$ and/or  $(\gdualn{\elko}\elko)^2$ would be generated by the flow. We have found that they are not generated, which is in agreement with the 1-loop renormalization study performed in ~\cite{deGracia:2024umr}~. This result could also be inferred on symmetry grounds, since the derivative Yukawa-like interaction has a shift-symmetry $\elko \mapsto \elko + \varepsilon$, where $\varepsilon$ is a constant ELKO spinor, which prevents the generation of interaction terms without derivative acting on $\elko$. 
	
	The set of renormalization group equations in this sub-section has a remarkable property: it allows UV-complete trajectories connecting the free fixed-point,
	\begin{equation}
		(m_\varphi^2, m_\elko^2, \lambda_\varphi, g_{\varphi\elko})_* =(0,0,0,0) \,,	
	\end{equation}
	in the deep UV, with finite couplings in the IR. This is possible thanks to the sign of $\beta_{g_{\varphi\elko}}$, which has a strictly negative coefficient multiplying $g_{\varphi\elko}^3$, therefore, allowing us to construct asymptotic-free trajectories for the couplings $g_{\varphi\elko}$. The beta function $\beta_{\lambda_\varphi}$ does not have a definite sign, however, we can still accommodate UV-complete trajectories connected with $\lambda_{\varphi*}=0$ since the term proportional to $g_{\varphi\elko}^4$ in $\beta_{\lambda_\varphi}$ generates non-vanishing (and positive) values of $\lambda_{\varphi}$ when we flow towards the IR. The UV-completion of the mass parameters is guaranteed by the negative canonical terms in the beta functions $\beta_{m_\varphi^2}$ and $\beta_{m_\elko^2}$. Note that $m_\elko^2 = 0$ is a partial fixed-point solution; therefore, from the renormalization group point of view, we can consistently set $m_\elko^2 = 0$ along all scales.
	
	In conclusion, the system discussed in this sub-section provides an example of UV-complete theory in 4-dimensions, which is perturbatively reliable since it is realized by a free fixed-point.
	
	\subsection{Scalar-ELKO portal and four-fermion self-interactions \label{subsec:ScalarELKO_2}}
	
	Now, we turn our attention to another version of a scalar-ELKO system, characterized by non-derivative interactions. The new interaction part of our truncation for the flowing action is given by
	\begin{equation}\label{eq:interaction_scalar+elko}
		\Gamma_k^{\text{int.}}[\elko,\gdualn{\elko},\varphi] = \int\!d^4x \,\left( \frac{1}{2} Z_\varphi Z_\elko \lambda_{\varphi\elko}\, \varphi^2 \gdualn{\elko}_a \elko_a + 
		Z_\elko^2 \lambda_{\elko} (\gdualn{\elko}_a \elko_a)^2
		+ \frac{1}{4!} Z_\varphi^2 \lambda_{\varphi}\, \varphi^4 \right) \,,
	\end{equation}
	where $\lambda_{\varphi\elko}$, $\lambda_{\elko}$ and $\lambda_{\varphi}$ are scale-dependent couplings. The first term in $\Gamma_k^{\text{int.}}[\elko,\gdualn{\elko},\varphi]$ is a scalar-ELKO portal. This term can be extended to a Higgs-portal by simply replacing $\varphi^2$ with $H^\dagger H$, where $H$ is the Higgs doublet. 
	In the proposals where ELKOs appear as dark-matter candidates, this would be one of the few perturbatively renormalizable interactions between the visible and dark sectors. The remaining two terms are self-interactions. As we will see, they are unavoidable once the scalar-ELKO portal is present.
	
	We compute the beta functions $\beta_{\lambda_{\varphi\elko}}$, $\beta_{\lambda_{\elko}}$ and $\beta_{\lambda_{\varphi}}$ by looking at the flow of the 4-point functions $\Gamma_{\varphi\varphi\elko\gdualn{\elko}}^{(4)}$ , $\Gamma_{\elko\elko\gdualn{\elko}\gdualn{\elko}}^{(4)}$and $\Gamma_{\varphi\varphi\varphi\varphi}^{(4)}$, respectively. Schematically, we can write
	\begin{subequations}
		\begin{equation}
			\beta_{\lambda_{\varphi\elko}} = (\eta_\elko + \eta_\varphi)\lambda_{\varphi\elko} + \mathcal{P}_{\lambda_{\varphi\elko}} \circ \left( k\partial_k \Gamma_{\varphi\varphi\elko\gdualn{\elko}}^{(4)}\right) \,,
		\end{equation}
		\begin{equation}
			\beta_{\lambda_{\elko}} = 2\,\eta_\elko \lambda_{\elko} + \mathcal{P}_{\lambda_{\elko}} \circ \left( k\partial_k \Gamma_{\elko\elko\gdualn{\elko}\gdualn{\elko}}^{(4)}\right) \,,
		\end{equation}
		\begin{equation}
			\beta_{\lambda_{\varphi}} =  2\,\eta_\varphi \lambda_{\varphi} + \mathcal{P}_{\lambda_{\varphi}} \circ \left( k \partial_k \Gamma_{\varphi\varphi\varphi\varphi}^{(4)}  \right)\,.
		\end{equation}
	\end{subequations}
	To properly define the projectors $\mathcal{P}_{\lambda_{\varphi\elko}}$ and $\mathcal{P}_{\lambda_{\elko}}$, we need to make sure that their application on the 4-point functions does not have any overlap with other tensor structures that could be generated by the renormalization group flow. At the level of our truncation, where $\Gamma_k^{\text{int.}}[\elko,\gdualn{\elko},\varphi]$ includes only non-derivative interactions, we need to project out tensor structures related to the following set of operators:
	\begin{equation}\label{eq:list_operator_fourfield}
		\begin{aligned}
			&\quad \varphi^2  \gdualn{\elko}_a \gamma_5 \elko_a \,,\quad 
			\gdualn{\elko}_a  \elko_a  \gdualn{\elko}_a \gamma_5 \elko_a  \,,\quad 
			\gdualn{\elko}_a \gamma_5 \elko_a \gdualn{\elko}_b \gamma_5 \elko_b \,,\quad
			\gdualn{\elko}_a \gamma_\mu \elko_a  \gdualn{\elko}_b \gamma^\mu \elko_b \,, \\
			&\gdualn{\elko}_a \gamma_\mu  \elko_a  \gdualn{\elko}_b \gamma^\mu \gamma_5 \elko_b \,,\quad 
			\gdualn{\elko}_a \gamma_\mu \gamma_5 \elko_a  \gdualn{\elko}_b \gamma^\mu \gamma_5 \elko_b \,,\quad 
			\gdualn{\elko}_a [\gamma_\mu,\gamma_\nu]  \elko_a  \gdualn{\elko}_b [\gamma^\mu,\gamma^\nu]  \elko_b  \,,
		\end{aligned}
	\end{equation}
	which, when combined with the interactions already present in \eqref{eq:interaction_scalar+elko}, defines a complete set of non-derivative four-field interactions compatible with the field content and symmetries of our system. Moreover, we can define a set of projectors associated with one of the operators in \eqref{eq:list_operator_fourfield}. By computing the flow of the 4-point functions $\Gamma_{\varphi\varphi\elko\gdualn{\elko}}^{(4)}$ and $\Gamma_{\elko\elko\gdualn{\elko}\gdualn{\elko}}^{(4)}$, and then acting with the list of projectors associated with \eqref{eq:list_operator_fourfield}, we can explicitly verify that none of these operators are generated by renormalization group. Therefore, we can safely ignore the list of operators \eqref{eq:list_operator_fourfield} in our truncation. On the structural level, we could already expect that none of these additional would be generated as they all involve gamma matrices, which are absent our original truncation.
	
	Using the flow of the 2-point functions $\Gamma_{\elko\gdualn{\elko}}^{(2)}$ and $\Gamma_{\varphi\varphi}^{(2)}$, we can compute the anomalous dimensions and the beta functions for the (dimensionless) mass parameters. In the present truncation, both anomalous dimensions vanish. For the flow of the mass parameters, we find the following results
	\begin{subequations}
		\begin{equation}
			\beta_{m_\varphi^2} = -2\,m_\varphi^2 - \frac{\lambda_{\varphi}}{32\pi^2 (1+m_\varphi^2)^2} + \frac{N_\text{E}\,\lambda_{\varphi\elko}}{4\pi^2 (1+m_\elko^2)^2} \,,
		\end{equation}
		\begin{equation}
			\beta_{m_\elko^2} = - 2\,m_\elko^2 + \frac{(8N_\text{E} -2)\, \lambda_{\elko}}{\pi^2 (1+m_\elko^2)^2} - \frac{\lambda_{\varphi\elko}}{2\pi^2 (1+m_\varphi^2)^2} \,.
		\end{equation}
	\end{subequations}
	Going back to the beta functions $\beta_{\lambda_\varphi}$, $\beta_{\lambda_{\varphi\elko}}$ and $\beta_{\lambda_{\elko}}$, the explicit results are
	\begin{subequations}
		\begin{equation}
			\beta_{\lambda_\varphi} = \frac{3 \lambda_{\varphi}^2}{16\pi^2 (1+m_\varphi^2)} - \frac{3N_\text{E}\, \lambda_{\varphi\elko}^2}{2\pi^2 (1+m_\elko^2)^3} \,,
		\end{equation}
		\begin{equation}
			\beta_{\lambda_{\varphi\elko}} = \frac{(2+m_\elko^2+m_\varphi^2)\,\lambda_{\varphi\elko}^2}{8\pi^2(1+m_\elko^2)^2(1+m_\varphi^2)^2}
			+ \frac{\lambda_\varphi \lambda_{\varphi\elko}}{16\pi^2(1+m_\varphi^2)^3}
			- \frac{(4N_\text{E}-1)\,\lambda_\elko \lambda_{\varphi\elko}}{4\pi^2(1+m_\elko^2)^3}\,,
		\end{equation}
		\begin{equation}
			\beta_{\lambda_{\elko}} = - \frac{(N_\text{E}-1) \lambda_{\elko}^2}{\pi^2 (1+m_\elko^2)^3} +  \frac{\lambda_{\varphi\elko}^2}{32\pi^2 (1+m_\varphi^2)^3} \,.
		\end{equation}
	\end{subequations}
	
	First, let us comment on the beta function $\beta_{\lambda_{\elko}}$ discarding the portal coupling $\lambda_{\varphi\elko}$. In this case, our result reduces to
	\begin{equation}
		\beta_{\lambda_{\elko}}\Big|_{\lambda_{\varphi\elko} = 0} = - \frac{(N_\text{E}-1) \lambda_{\elko}^2}{\pi^2 (1+m_\elko^2)^3} \,.
	\end{equation}
	This result should be compared with ~\cite{deBrito:2019hih}~, where we have found that the beta function for $\lambda_{\elko}$ vanishes at 1-loop. The analysis done in ~\cite{deBrito:2019hih} was restricted to the case $N_\text{E} = 1$; therefore, the result presented in this paper shows that the vanishing beta function reported in ~\cite{deBrito:2019hih} does not survive if we consider a system with more than one ELKO. Instead, if we consider $N_\text{E} > 1$, then the four-fermion coupling $\lambda_{\elko}$ becomes asymptotically free (assuming $\lambda_{\elko}>0$ at finite scales). However, it does not imply UV-completion of the full system. If one sets $\lambda_{\varphi\elko} = 0$, then the beta function for the quartic scalar coupling $\lambda_\varphi$ becomes strictly positive and, therefore, we are back to the usual Landau-pole/triviality problem of $\varphi^4$-theories.
	
	Now, going back to the full system of renormalization group equations, we investigate whether this version of the scalar-ELKO system is compatible with UV completion. Once again, we focus in the possibility of UV-complete trajectories associated with the free fixed-point 
	\begin{equation}
		(m_\varphi^2,m_\elko^2,\lambda_{\varphi},\lambda_{\varphi\elko},\lambda_{\elko})_* = (0,0,0,0,0).
	\end{equation}
	As one can see, in each one of the beta functions, there is a competition between screening (positive) and anti-screening (negative) contributions. 
	Therefore, we can expect that not all points in the coupling space can be connect with the free-fixed point via renormalization group trajectory. 
	
	We can obtain more information by restricting our analysis to a region infinitesimally close to the free fixed-point. Since this region is supposed realized in the deep UV regime, we can ignore the mass parameters $m_\varphi^2$ and $m_\elko^2$, and focus on the beta functions $\beta_{\lambda_\varphi}$, $\beta_{\lambda_{\varphi\elko}}$ and $\beta_{\lambda_{\elko}}$ with $m_\varphi^2=0$ and $m_\elko^2=0$.
	In this case, we can write
	\begin{subequations}
		\begin{equation}
			\beta_{\lambda_\varphi} = \frac{3}{16\pi^2} \left(\lambda_{\varphi}^2 - 8N_\text{E}\lambda_{\varphi\elko}^2\right) \,,
		\end{equation}
		\begin{equation}
			\beta_{\lambda_{\varphi\elko}} = \frac{1}{16\pi^2} \left( 4 \lambda_{\varphi\elko}^2 + \lambda_\varphi \lambda_{\varphi\elko} - 4(4N_\text{E}-1)\lambda_{\elko} \lambda_{\varphi\elko} \right)\,,
		\end{equation}
		\begin{equation}
			\beta_{\lambda_{\elko}} = - \frac{1}{32\pi^2} \left(  32\,(N_\text{E}-1) \lambda_{\elko}^2 - \lambda_{\varphi\elko}^2  \right) \,.
		\end{equation}
	\end{subequations}
	Since the flow is quadratic in all couplings, we cannot investigate the stability properties of the free fixed-point in terms of linearized flows. Instead, we consider the strategy discussed in the \ref{app:LO-Flow} to integrate the flow in an infinitesimal neighborhood around the free fixed-point.
	In general, the qualitative results depends on whether we are consider $N_\text{E} =1$ or $N_\text{E} >1$, therefore, we will separate the discussion into two cases.
	
	\subsubsection{Leading order flow for $N_\text{E}=1$}
	
	In this case, we were able to find three types of solutions compatible with the free-fixed point in the UV. They are given by
	\begin{subequations}
		\begin{equation}\label{eq:Scalar+ELKO-L0-Flow_1_N=1}
			(\lambda_\varphi(k),\lambda_{\varphi\elko}(k),\lambda_{\elko}(k)) \approx C_0 \big(-52.64 \,, \, 0 \, , \, 0 \, \big) \times  \big[\log(k/k_0) \big]^{-1} \,,
		\end{equation}
		\begin{equation}
			\hspace*{-.1cm}
			\begin{aligned}\label{eq:Scalar+ELKO-L0-Flow_2_N=1}
				(\lambda_\varphi(k),\lambda_{\varphi\elko}(k),\lambda_{\elko}(k)) 
				& \approx C_0 \, \big(-473.74 \,, \, -157.91 \, , \, -78.96 \, \big) \times  \big[\log(k/k_0) \big]^{-1} \\
				&\hspace*{-.35cm} +C_1 \, \big(-123.68 \,,\, +240.49  \, , \, +261.11 \, \big) \times  \big[\log(k/k_0) \big]^{-9} \\
				&\hspace*{-.35cm} +C_2 \, \big(+154.33 \,,\, -332.04 \, , \, -261.89  \, \big) \times  \big[\log(k/k_0) \big]^{-12}  \,,
			\end{aligned}
		\end{equation}
		\begin{equation}\label{eq:Scalar+ELKO-L0-Flow_3_N=1}
			\begin{aligned}
				(\lambda_\varphi(k),\lambda_{\varphi\elko}(k),\lambda_{\elko}(k)) 
				& \approx C_0 \,\big(-90.97 \,,\, -20.88 \, , \, -1.38 \, \big) \times  \big[\log(k/k_0) \big]^{-1} \\
				& +C_1 \,\big(+20.88 \,,\,-90.50 \, , \, -7.30 \, \big) \times  \big[\log(k/k_0) \big]^{-2.86}   \,,
			\end{aligned}
		\end{equation}
	\end{subequations}
	where $C_0 (\geq 0)$, $C_1$ and $C_2$ are free parameters that appears as integration constants, and $k_0$ is a reference scale. 
	
	The first type of solution (Eq. \eqref{eq:Scalar+ELKO-L0-Flow_1_N=1}) enforces $\lambda_{\varphi\elko} = \lambda_{\elko} = 0$ along the flow. In this case, we are back to the flow of a standard $\varphi^4$-theory, where it is well known that we cannot construct non-trivial solutions with positive couplings in the IR. This is reflected on the negative sign of $\lambda_\varphi(k)$ in \eqref{eq:Scalar+ELKO-L0-Flow_1_N=1}. Therefore, this solution is not physical acceptable.
	
	The second and third type of solutions (Eqs. \eqref{eq:Scalar+ELKO-L0-Flow_2_N=1} and \eqref{eq:Scalar+ELKO-L0-Flow_3_N=1}) are compatible with non-vanishing values for all couplings. However, it is likely that they are not compatible with positive values of $\lambda_\varphi(k)$ in the IR.
	Unfortunately, Eqs. \eqref{eq:Scalar+ELKO-L0-Flow_2_N=1} and \eqref{eq:Scalar+ELKO-L0-Flow_3_N=1} are not sufficient for us to understand if the flow is really incompatible with positive values of $\lambda_\varphi(k)$ in the IR, since higher-order terms that are not included in our approximation might play an important role in the flow. 
	
	\subsubsection{Leading order flow for $N_\text{E}\geq 2$}
	
	In this case, we were able to find four types of solutions compatible with the free-fixed point in the UV. Focusing on $N_\text{E} = 2$ as a concrete example, we find
	\begin{subequations}
		\begin{equation}\label{eq:Scalar+ELKO-L0-Flow_1_N=2}
			(\lambda_\varphi(k),\lambda_{\varphi\elko}(k),\lambda_{\elko}(k)) \approx C_0 \,\big(-52.64 \,, \, 0 \, , \, 0 \, \big) \times  \big[\log(k/k_0) \big]^{-1} \,,
		\end{equation}
		\begin{equation}
			\begin{aligned}\label{eq:Scalar+ELKO-L0-Flow_2_N=2}
				(\lambda_\varphi(k),\lambda_{\varphi\elko}(k),\lambda_{\elko}(k)) 
				& \approx C_0 \, \big( -52.64 \,, \, 0 \, , \, +9.87 \, \big) \times  \big[\log(k/k_0) \big]^{-1} \\
				& +C_1 \,\big( +9.87 \,, \, 0 \, , \, +52.64 \, \big) \times  \big[\log(k/k_0) \big]^{-2} \\
				& +C_2 \,\big( 0 \,, \, -9.87 \, , \, 0 \, \big)  \times  \big[\log(k/k_0) \big]^{-2.08}  \,,
			\end{aligned}
		\end{equation}
		\begin{equation}
			\hspace*{-.05cm}
			\begin{aligned}\label{eq:Scalar+ELKO-L0-Flow_2_N=2}
				(\lambda_\varphi(k),\lambda_{\varphi\elko}(k),\lambda_{\elko}(k)) 
				& \approx C_0 \, \big( -109.90 \,, \, -19.83 \, , \, -1.12 \, \big) \times  \big[\log(k/k_0) \big]^{-1} \\
				& +C_1 \,\big( +19.82 \,, \, -109.43 \, , \, -7.45 \, \big) \times  \big[\log(k/k_0) \big]^{-3.44}  \,,
			\end{aligned}
		\end{equation}
		\begin{equation}\label{eq:Scalar+ELKO-L0-Flow_3_N=2}
			\begin{aligned}
				(\lambda_\varphi(k),\lambda_{\varphi\elko}(k),\lambda_{\elko}(k)) 
				& \approx C_0 \,\big( 0 \,, \, +9.87 \, , \, 0 \, \big) \times  \big[\log(k/k_0) \big]^{-1} \\
				& +C_1 \,\big( 0 \,, \, 0 \, , \, +9.87 \, \big)  \times  \big[\log(k/k_0) \big]^{-1.75}   \,,
			\end{aligned}
		\end{equation}
	\end{subequations}
	where $C_0 (\geq 0)$, $C_1$ and $C_2$ are free-parameters, and $k_0$ is a reference scale.
	
	The numerical coefficients and exponents appearing in this set of solutions are specific to $N_\text{E} = 2$. However, we have explicitly verified that the qualitative aspects of these solutions are also valid for $N_E>2$.
	
	The first, second and third solutions are qualitatively similar to the ones obtained with $N_\text{E} = 1$, therefore, the previous discussion also applies to the case of $N_{E} \geq 2$.
	
	The fourth solution, Eq. \eqref{eq:Scalar+ELKO-L0-Flow_3_N=2}, has a richer structure. In this case, from the leading-order solution around the free fixed-point, we can see that it is possible to construct asymptotically-free trajectories with non-vanishing values of $\lambda_{\varphi\elko}$ and $\lambda_{\varphi}$ away from the fixed-point. Moreover, once the portal coupling $\lambda_{\varphi\elko}$ departs from the fixed-point regime, the beta function $\lambda_{\varphi}$ implies the generation of the quartic coupling $\lambda_{\varphi}$ with positive values in the IR. Therefore, we can conclude that this system is compatible with asymptotically free solutions. 
	
	Once we depart further away from the fixed-point regime, we need to take into account higher-order effects, and the solutions discussed here are no longer a good approximation. Such higher-order effects might lead to non-trivial restrictions on the IR values of the couplings. However, this is beyond the scope of our analysis.
	
	\section{ELKO interacting with a gauge field via Pauli-like interaction \label{sec:ELKO+Gauge}}
	
	In this section, we investigate the Pauli-like interaction term involving ELKOs and an Abelian gauge field. 
	To simplify the discussion, we will not add any interaction involving the scalar field; therefore, we can just neglect $\Gamma_k^{\text{scalar}}[\varphi]$ in our truncation.
	
	The goal of this section is to answer the following question: i) What is the minimal set of (perturbatively renormalizable) operators that is necessary when the Pauli-like interaction is present? ii) Is the minimal gauge-ELKO system compatible with UV completion?
	
	\subsection{The necessity of extra four-fermion channels \label{subsec:PauliInduced4Ferm}}
	
	To answer the first question, we set up the following interaction term
	\begin{equation}
		\begin{aligned}
			\Gamma_k^{\text{int.}}[\elko,\gdualn{\elko},A] &=  
			\int\!\! d^4x  \left( \frac{1}{2} Z_A^{1/2} Z_\elko \, g_{A\elko} \, F_{\mu\nu} \gdualn{\elko} \,[\gamma^\mu,\gamma^\nu] \,\elko \right)+ \\
			&+  
			\sum_{n=1}^{7}  \int\!\! d^4x  \left( Z_\elko^2 \, \lambda_{\elko}^{(n)}\, \mathcal{O}_{n}^{\text{4-ELKO}}[\elko,\gdualn{\elko}]\right) \,,
		\end{aligned}
	\end{equation}
	where $g_{A,\elko}$ and $\lambda_{\elko}^{(n)}$'s are scale-dependent couplings, and $\{\mathcal{O}_{n}^{\text{4-ELKO}}\}_{n=1,\cdots,7}$ defines a complete set of non-derivative four-fermion operators constructed with ELKOs. The explicit four-fermion operators are given by
	\begin{subequations}
		\begin{equation}
			\mathcal{O}_{1}^{\text{4-ELKO}} = \gdualn{\elko}_a \elko_a \, \gdualn{\elko}_b \elko_b \,,
		\end{equation}
		\begin{equation}
			\mathcal{O}_{2}^{\text{4-ELKO}} =  \gdualn{\elko}_a \elko_a \, \gdualn{\elko}_b \gamma_5 \elko_b \,,
		\end{equation}
		\begin{equation}
			\mathcal{O}_{3}^{\text{4-ELKO}} = \gdualn{\elko}_a \gamma_5 \elko_a \, \gdualn{\elko}_b \gamma_5 \elko_b \,,
		\end{equation}
		\begin{equation}
			\mathcal{O}_{4}^{\text{4-ELKO}} = \gdualn{\elko}_a \gamma_\mu \elko_a \, \gdualn{\elko}_b \gamma^\mu \elko_b \,,
		\end{equation}
		\begin{equation}
			\mathcal{O}_{5}^{\text{4-ELKO}} =\gdualn{\elko}_a \gamma_\mu  \elko_a \, \gdualn{\elko}_b \gamma^\mu\gamma_5 \elko_b \,,
		\end{equation}
		\begin{equation}
			\mathcal{O}_{6}^{\text{4-ELKO}} =\gdualn{\elko}_a \gamma_\mu \gamma_5 \elko_a \, \gdualn{\elko}_b \gamma^\mu\gamma_5 \elko_b \,,
		\end{equation}
		\begin{equation}
			\mathcal{O}_{7}^{\text{4-ELKO}} =\frac{1}{4}\gdualn{\elko}_a [\gamma_\mu,\gamma_\nu] \elko_a \, \gdualn{\elko}_b [\gamma^\mu,\gamma^\nu] \elko_b \,.
		\end{equation}
	\end{subequations}
	
	We start by looking at the renormalization group flow of the four-fermion interactions. In general, the schematic structure of the beta functions is given by
	\begin{equation}\label{eq:generalfourfermbeta}
		\beta_{\lambda_{\elko}^{(n)}} = 2 \, \eta_\elko \lambda_{\elko}^{(n)} + \mathcal{P}_{\lambda_{\elko}^{(n)}} \circ \left( k \partial_k \Gamma_{\elko\elko\gdualn{\elko}\gdualn{\elko}}^{(4)}  \right) \,.
	\end{equation}
	
	First, we check which four-fermion interactions are directly generated by the Pauli-like interaction. This can be done by computing the second term in the right-hand side of \eqref{eq:generalfourfermbeta}, and then setting $\lambda_{\elko}^{(n)} = 0$ for all values of $n$. The resulting projected flows are
	\begin{subequations}
		\begin{equation}
			\left[ \mathcal{P}_{\lambda_{\elko}^{(1)}} \circ \left( k \partial_k \Gamma_{\elko\elko\gdualn{\elko}\gdualn{\elko}}^{(4)}  \right) \right]_{\lambda_\elko^{(m)}=0} = \frac{3\,(2+m_\elko^2)\,g_{A\elko}^4}{\pi^2 (1+m_\eta^2)^3} \,,
		\end{equation}
		\begin{equation}
			\left[ \mathcal{P}_{\lambda_{\elko}^{(n)}} \circ \left( k \partial_k \Gamma_{\elko\elko\gdualn{\elko}\gdualn{\elko}}^{(4)}  \right) \right]_{\lambda_\elko^{(m)}=0} = 0 \,,\qquad (n\neq1) \,.
		\end{equation}
	\end{subequations}
	Therefore, $(\gdualn{\elko}_a\elko_a)^2$ is the only four-fermion interaction directly generated by the Pauli-like term.
	
	Next, we investigate what interactions are indirectly generated by the Pauli-like term. Since we established that $\mathcal{O}_{1}^{\text{4-ELKO}} = (\gdualn{\elko}_a\elko_a)^2$ is generated by the Pauli-like term, a consistent flow requires $\lambda_{\elko}^{(1)} \neq 0$. This will feed back into the flow of $\Gamma_{\elko\elko\gdualn{\elko}\gdualn{\elko}}^{(4)} $ and possibly generate other four-fermion interactions. We have explicitly checked that:
	\begin{itemize}
		\item[i)] Setting $\lambda_{\elko}^{(1)} \neq 0$, the renormalization group flow generates $\mathcal{O}_{7}^{\text{4-ELKO}} \sim (\gdualn{\elko}_a \,[\gamma_\mu,\gamma_\nu] \,\elko_a)^2$, therefore demanding $\lambda_{\elko}^{(7)} \neq0$.
		\item[ii)] Setting $\lambda_{\elko}^{(1)} \neq 0$ and $\lambda_{\elko}^{(7)} \neq 0$, the renormalization group flow generates $\mathcal{O}_{3}^{\text{4-ELKO}} \sim (\gdualn{\elko}_a \,\gamma_5 \,\elko_a)^2$, therefore, demanding $\lambda_{\elko}^{(3)} \neq0$.
		\item[iii)] Setting $\lambda_{\elko}^{(1)} \neq 0$,  $\lambda_{\elko}^{(3)} \neq 0$, and $\lambda_{\elko}^{(7)} \neq 0$, the renormalization group flow does not generate any additional four-fermion interaction.
	\end{itemize}
	Therefore, the minimal set of four-fermion interactions that are necessary for a self-consistent flow is given by $\{\mathcal{O}_{1}^{\text{4-ELKO}},\mathcal{O}_{3}^{\text{4-ELKO}},\mathcal{O}_{7}^{\text{4-ELKO}}\}$.
	
	Since we have established that the operators $\mathcal{O}_{2}^{\text{4-ELKO}}$, $\mathcal{O}_{4}^{\text{4-ELKO}}$, $\mathcal{O}_{5}^{\text{4-ELKO}}$ and $\mathcal{O}_{6}^{\text{4-ELKO}}$ are not mandatory for a self-consistent flow, we can consistently set $\lambda_{\elko}^{(2)} = \lambda_{\elko}^{(4)} = \lambda_{\elko}^{(5)} = \lambda_{\elko}^{(6)} = 0$ in our truncation. However, if we consider $\lambda_\elko^{(2)} \neq 0$, the flow of $\Gamma_{\elko\gdualn{\elko}}^{(2)}$ and $\Gamma_{A\elko\gdualn{\elko}}^{(3)}$ would generate two new operators given by
	\begin{equation}
		\mathcal{O}_{\elko\gdualn{\elko}} \sim  \gdualn{\elko}_a \gamma_5 \elko_a
		\qquad \text{and} \qquad
		\mathcal{O}_{A\elko\gdualn{\elko}} \sim  \tilde{F}_{\mu\nu} \gdualn{\elko}_a \,[\gamma^\mu,\gamma^\nu] \,\elko_a \,,
	\end{equation}
	where $\tilde{F}_{\mu\nu} = \frac{1}{2} \epsilon_{\mu\nu\alpha\beta} F^{\alpha\beta}$ is the dual of the field-strength $F_{\mu\nu}$. In this case, we would need to include these new operators in our truncation, which potentially generates further four-fermion interactions.
	
	\subsection{The possibility of asymptotically free trajectories \label{subsec:AFGaugeELKO}}
	
	Once established the minimal set of operators necessary for a self-consistent truncation of a gauge-ELKO system, we now investigate the corresponding flow equations in such a minimal setting.
	
	We start with the flow of the Pauli-like coupling $g_{A\elko}$. Schematically, the corresponding beta functions is given by
	\begin{equation}
		\beta_{g_{A\elko}} = \left( \frac{1}{2} \eta_A + \eta_\elko \right) g_{A\elko}
		+ \mathcal{P}_{g_{A\elko}} \circ \left(k\partial_k \Gamma_{A\elko\gdualn{\elko}}^{(3)}\right) \,.
	\end{equation}
	Using the flow of the 2-point functions $\Gamma_{AA}^{(2)}$ and $\Gamma_{\elko\gdualn{\elko}}^{(2)}$ to evaluate the anomalous dimensions, we get
	\begin{subequations}
		\begin{equation}
			\eta_A = - \frac{2\,N_\text{E} \, g_{A\elko}^2}{\pi^2 (1+m_\elko^2)} \,,
		\end{equation}
		\begin{equation}
			\eta_\elko = 0 \,.
		\end{equation}
	\end{subequations}
	For the (dimensionless) mass parameter $m_\elko^2$, we find the
	\begin{equation}
		\beta_{m_\elko^2} = -2 m_\elko^2 - \frac{8\, (2+m_\elko^2) g_{A\elko}^2}{\pi^2 (1+m_\elko^2)^2} + \frac{(8N_\text{E}-2) \lambda_{\elko}^{(1)} -2\,\lambda_{\elko}^{(3)} +24\,\lambda_{\elko}^{(7)} }{\pi^2 (1+m_\elko^2)^2} \,.
	\end{equation}
	Going back to $\beta_{g_{A\elko}}$, we find the following expression
	\begin{equation}
		\beta_{g_{A\elko}} = - \frac{(3+6\,N_\text{E}+m_\elko^2)\,g_{A\elko}^3}{6\pi^2 (1+m_\elko^2)^3}
		+ \frac{\big(\lambda_{\elko}^{(1)} + \lambda_{\elko}^{(3)} + (4+8\,N_\text{E}) \, \lambda_{\elko}^{(7)}\big)\,g_{A\elko}}{4\pi^2 (1+m_\elko^2)^3} \,.
	\end{equation}
	The beta function of the Pauli-like terms was previously computed in ~\cite{Carvalho:2023btn} using the minimal subtraction scheme. However, our result has a few differences in comparison with ~\cite{Carvalho:2023btn}~: 
	i) the appearance of mass thresholds is a general feature of the FRG which is absent in the minimal subtraction scheme; 
	ii) the analysis performed in ~\cite{Carvalho:2023btn} was restricted to $N_\text{E} = 1$ and did not include the couplings $\lambda_{\varphi}^{(3)}$ and $\lambda_{\varphi}^{(7)}$; 
	iii) we used different normalization factors for the couplings, leading to different factors in the beta functions. 
	Nevertheless, our result agrees with one of the main features observed in ~\cite{Carvalho:2023btn}~, the $g_{A\elko}^3$ is anti-screening (negative), while the contributions involving the four-fermion couplings are screening (positive).
	
	Similarly to what is usually done in gauge theories, it is convenient to work with the beta function of $g_{A\elko}^2$ instead of $g_{A\elko}$. The former is related with the later according to $\beta_{g_{A\elko}^2} = 2 \,g_{A\elko}\,\beta_{g_{A\elko}}$. Therefore, we can write
	\begin{equation}
		\beta_{g_{A\elko}^2} = - \frac{(3+6\,N_\text{E}+m_\elko^2)\,g_{A\elko}^4}{3\pi^2 (1+m_\elko^2)^3}
		+ \frac{\big(\lambda_{\elko}^{(1)} + \lambda_{\elko}^{(3)} + (4+8\,N_\text{E}) \, \lambda_{\elko}^{(7)}\big)\,g_{A\elko}^2}{2\pi^2 (1+m_\elko^2)^3} \,.
	\end{equation}
	
	For the four-fermion couplings $\lambda_{\elko}^{(1)}$, $\lambda_{\elko}^{(3)}$ and $\lambda_{\elko}^{(7)}$, we find the following set of beta functions
	\begin{subequations}
		\begin{equation}
			\begin{aligned}
				\beta_{\lambda_\elko^{(1)}} &= 
				\frac{-2\,(N_\text{E}-1)(\lambda_{\elko}^{(1)})^2 + (\lambda_{\elko}^{(3)})^2 + 24(\lambda_{\elko}^{(7)})^2 + \lambda_{\elko}^{(1)} \lambda_{\elko}^{(3)}- 12\lambda_{\elko}^{(1)}\lambda_{\elko}^{(7)}}{2\pi^2 (1+m_\elko^2)^3} \\
				&+ \frac{(3+m_\elko^2) (\lambda_{\elko}^{(1)} - 4 \lambda_{\elko}^{(7)}) g_{A\elko}^2}{\pi^2 (1+m_\elko^2)^3} + \frac{3\,(2+m_\elko^2)\,g_{A\elko}^4}{\pi^2 (1+m_\eta^2)^3} \,,
			\end{aligned}
		\end{equation}
		\begin{equation}
			\begin{aligned}
				\beta_{\lambda_\elko^{(3)}} &= 
				\frac{- (2N_\text{E}-1) (\lambda_{\elko}^{(3)})^2 + 24 (\lambda_{\elko}^{(7)})^2 + 3\lambda_{\elko}^{(1)}\lambda_{\elko}^{(3)}- 12 \lambda_{\elko}^{(3)} \lambda_{\elko}^{(7)}}{2\pi^2(1+m_\elko^2)^3} \\
				&+ \frac{(3+m_\elko^2) (\lambda_{\varphi}^{(3)} - 4 \lambda_{\varphi}^{(7)}) g_{A\elko}^2}{\pi^2 (1+m_\elko^2)^3} \,,
			\end{aligned}
		\end{equation}
		\begin{equation}\hspace{-1cm}
			\begin{aligned}
				\beta_{\lambda_\elko^{(7)}} &= 
				\frac{4\,(N_\text{E}+1) (\lambda_{\elko}^{(7)})^2 + 3\, \lambda_{\elko}^{(1)} \lambda_{\elko}^{(7)} + 3\, \lambda_{\elko}^{(3)} \lambda_{\elko}^{(7)}}{2\pi^2(1+m_\elko^2)^3} \\
				&- \frac{(3+m_\elko^2) ( \lambda_{\varphi}^{(1)} + \lambda_{\varphi}^{(3)} + 2 \lambda_{\varphi}^{(7)}) g_{A\elko}^2}{\pi^2 (1+m_\elko^2)^3} \,.
			\end{aligned}
		\end{equation}
	\end{subequations}

	Now, we investigate the possibility of constructing UV-complete renormalization group trajectories connecting non-vanishing couplings in the IR with the free fixed-point,
	\begin{equation}
		(m_\eta^2,g_{A\elko},\lambda_{\varphi\elko}^{(1)},\lambda_{\varphi\elko}^{(3)},\lambda_{\varphi\elko}^{(7)}) = (0,0,0,0,0) \,,
	\end{equation}
	in the deep UV regime.
	Similarly to the system discussed in the previous section, the viability of asymptotically free trajectories relies on a non-trivial balance between screening (positive sign) and anti-screening (negative sign) contributions on the beta functions. To obtain more information about this possibility, we investigate the leading order flow around the free fixed-point based on the strategy discussed in the \ref{app:LO-Flow}. Once again, we set the mass parameter $m_\elko^2 = 0$, thus focusing exclusively on the beta functions $\beta_{g_{A\elko}^2}$, $\beta_{\lambda_{\elko}^{(1)}}$, $\beta_{\lambda_{\elko}^{(3)}}$ and $\beta_{\lambda_{\elko}^{(7)}}$.
	
	First, let us focus on $N_\text{E}=1$. In this case, the leading order flow leads to two types of trajectories
	\begin{subequations}
		\begin{equation}\label{eq:Gauge+ELKO-L0-Flow_1_N=1}
			\begin{aligned}
				&(g_{A\elko}^2(k),\lambda_{\elko}^{(1)}(k),\lambda_{\elko}^{(3)}(k),\lambda_{\elko}^{(7)}(k)) \approx \\
				&\qquad\qquad\approx \,C_0 \,\big( +5.70\,,\, -7.79 \,,\, +11.95 \,,\, +0.86 \big) \times  \big[\log(k/k_0) \big]^{-1} \\
				&\qquad\qquad\,+ \,C_1 \,\big( -10.43 \,,\, +5.88 \,,\, +8.81 \,,\, +0.17  \big) \times  \big[\log(k/k_0) \big]^{-3} \,,
			\end{aligned}
		\end{equation}
		\begin{equation}\label{eq:Gauge+ELKO-L0-Flow_2_N=1}
			\begin{aligned}
				&(g_{A\elko}^2(k),\lambda_{\elko}^{(1)}(k),\lambda_{\elko}^{(3)}(k),\lambda_{\elko}^{(7)}(k)) \approx \\
				&\qquad\qquad\approx \,C_0 \,\big( +5.70 \,,\, -23.23 \,,\, -3.49 \,,\, +3.43 \big) \times  \big[\log(k/k_0) \big]^{-1} \\
				&\qquad\qquad\,+ \,C_1 \,\big( +0.19 \,,\, +1.02 \,,\, -6.66 \,,\, -0.18 \big) \times  \big[\log(k/k_0) \big]^{-3} \\
				&\qquad\qquad\,+ \,C_2 \,\big( -2.14 \,,\, +0.52 \,,\, -6.88 \,,\, +0.08 \big) \times  \big[\log(k/k_0) \big]^{-4.65} \,,
			\end{aligned}
		\end{equation}
	\end{subequations}
	where $C_0 (\geq 0)$, $C_1$ and $C_2$ are free parameters that appear as integration constants, and $k_0$ is a reference scale. 
	
	For $N_\text{E} \geq 2$, the leading order analysis of our system of flow equations reveals three types of solutions. As an example, we consider $N_\text{E} = 2$, for which we find the following trajectories
	\begin{subequations}
		\begin{equation}\label{eq:Gauge+ELKO-L0-Flow_1_N=2}
			\begin{aligned}
				&(g_{A\elko}^2(k),\lambda_{\elko}^{(1)}(k),\lambda_{\elko}^{(3)}(k),\lambda_{\elko}^{(7)}(k)) \approx \\
				&\qquad\qquad\approx \,C_0 \,\big( +9.80 \,,\, -30.03 \,,\, -22.32 \,,\, +6.53 \big) \times  \big[\log(k/k_0) \big]^{-1} \\
				&\qquad\qquad\,+ \,C_1 \,\big( -31.48 \,,\, +5.31 \,,\, -20.89 \,,\, +0.22 \big) \times  \big[\log(k/k_0) \big]^{-9.06} \,,
			\end{aligned}
		\end{equation}
		\begin{equation}\label{eq:Gauge+ELKO-L0-Flow_2_N=2}
			\begin{aligned}
				&(g_{A\elko}^2(k),\lambda_{\elko}^{(1)}(k),\lambda_{\elko}^{(3)}(k),\lambda_{\elko}^{(7)}(k)) \approx \\
				&\qquad\qquad\approx \,C_0 \,\big( 21.67 \,,\, +68.57 \,,\, +8.744 \,,\, +5.980 \big) \times  \big[\log(k/k_0) \big]^{-1} \\
				&\qquad\qquad\,+ \,C_1 \,\big( -54.63 \,,\, +21.47 \,,\, -25.13, -11.59 \big) \times  \big[\log(k/k_0) \big]^{-16.20} \,,
			\end{aligned}
		\end{equation}
		\begin{equation}\label{eq:Gauge+ELKO-L0-Flow_3_N=2}
			\begin{aligned}
				&(g_{A\elko}^2(k),\lambda_{\elko}^{(1)}(k),\lambda_{\elko}^{(3)}(k),\lambda_{\elko}^{(7)}(k)) \approx \\
				&\qquad\qquad\approx \,C_0 \,\big( +3.19 \,,\, -3.04 \,,\, 6.94 \,,\, 0.41 \big) \times  \big[\log(k/k_0) \big]^{-1} \\
				&\qquad\qquad\,+ \,C_1 \,\big( -6.18 \,,\, +2.96 \,,\, +4.15 \,,\, -0.2406 \big) \times  \big[\log(k/k_0) \big]^{-2.772} \,.
			\end{aligned}
		\end{equation}
	\end{subequations}
	Once again, $C_0 (\geq 0)$, $C_1$ and $C_2$ are free parameters arising as integration constants, and $k_0$ is a reference scale. We have checked that the qualitative features of these solutions remain unchanged for larger values of $N_\text{E}$. 
	
	At this point, the only restriction that we can impose on the flow is that $g_{A\elko}^2(k)$ should be positive for all values of $k$ to avoid complex-valued couplings. We can accommodate this restriction at the level of the leading-order flow by properly choosing the integration constants. Therefore, the leading order analysis suggests that the gauge-ELKO system can be compatible with asymptotically free trajectories.
	
	However, it is still unclear what are the restrictions on the IR values of the couplings that are compatible with asymptotically free solutions beyond the leading order analysis. This is beyond the scope of this paper.

	\section{Final remarks \label{sec:Conclusion}}
	
	In this paper, we investigated the structure of interactions involving mass dimension one fermions described by ELKO spinors from the perspective of the Wilsonian renormalization group. We used the functional renormalization group as a method to derive flow equations for interaction couplings. We investigated two types of scalar-ELKO systems, the first involving a derivative Yukawa-like interaction and the second involving a scalar-ELKO portal as well as four-fermion self-interactions. We also investigate gauge-ELKO system with a portal interaction given by a Pauli-like term.\vspace{.1cm}
	
	The main findings of this paper are the following:\vspace{.1cm}
	
	\begin{itemize}
		\item \textit{Scalar-ELKO system with a derivative Yukawa-like interaction:}\\
		We have explicitly verified that the renormalization group flow does not generate non-derivative four-fermion interactions. This property can be traced back to a softly-broken shift-symmetry that prevents the generation of non-derivative interactions due to UV divergences. This is aligned with the renormalization properties studied in ~\cite{deGracia:2024umr}~.\\
		We also discovered that this system has a remarkable property of being UV-complete due to the property of asymptotic freedom.\\
		It is unclear, however, how to make this system compatible with the symmetric phase of the Standard Model, since the derivative Yukawa coupling cannot be readily extended to accommodate the Higgs SU(2)-doublet.\vspace{.2cm}
		\item \textit{Scalar-ELKO system with scalar portal and four-fermion interactions}\\
		We found that the scalar-ELKO system including a single four-fermion term, namely $(\gdualn{\elko}_a\elko_a)^2$, does not generate further four-fermion interactions. This result is consistent with previous findings based on standard 1-loop renormalization ~\cite{deBrito:2019hih}~. However, we discovered one aspect that was overlooked in previous studies: if $N_\text{E} > 1$, then the beta function of the four-fermion couplings does not vanish as it was reported in ~\cite{deBrito:2019hih}~.\\
		If we neglect the scalar sector, then the pure ELKO system with $N_\text{E} > 1$ becomes asymptotically free. Once we include the interactions with scalars, the situation is more subtle, given that the system of beta functions shows a competition between screening (positive sign) and anti-screening (negative sign) contributions. Our analysis of the leading order flow around the free fixed-point reveals that one can accommodate asymptotically free trajectories, although it is still not clear what are the restrictions to the IR values of the couplings. \vspace{.2cm}
		\item \textit{Gauge-ELKO system with Pauli-like interaction}:\\
		We discovered that once the Pauli-like interaction is present, the renormalization group flow generates three four-fermion channels of ELKO self-interactions, two of them being so far unexplored in the literature. In contrast, there are four additional non-derivative four-fermion interactions that are not generated by the renormalization group flow.\\
		Performing a leading-order analysis of the flow, we found the possibility of accommodating free trajectories in the presence of the Pauli-like coupling. However, it is still unclear whether this conclusion remains stable beyond the leading order analysis, \textit{i.e.}, once we allow the couplings to flow further away from the fixed-point regime.
	\end{itemize}
	
	 
	Our results also open new directions of investigation. First, for the systems investigated in \ref{subsec:ScalarELKO_2} and \ref{subsec:AFGaugeELKO}, it would be interesting to understand what IR values of the couplings are compatible with asymptotically free trajectories once the flow departs from the neighborhood of the free fixed-point. This might lead to non-trivial restrictions on the interaction couplings. 
	
	Second, the anti-screening contribution of the $\lambda_{\varphi\elko}^2$-term in the beta function $\beta_{\lambda_\varphi}$ might have impact on questions related with Higgs (in)stability, potentially leading to non-trivial constraints on values of the portal coupling $\lambda_{\varphi\elko}$.
	
	Finally, the FRG has been used as a primary tool to study the asymptotic safety scenario for quantum gravity ~\cite{Weinberg:1980gg,Reuter:1996cp}~, see for instance ~\cite{Dupuis:2020fhh,Bonanno:2020bil,Knorr:2022dsx,Eichhorn:2022gku,Morris:2022btf,Saueressig:2023irs,Pawlowski:2023gym,Reichert:2020mja,Basile:2024oms,Percacci:2017fkn,Reuter:2019byg} for reviews, lecture notes, and textbooks on this topic. The present paper paves the way to future investigation of the asymptotically safe gravity-matter systems, including the ELKO as a dark matter candidate. 
	
	\section*{Acknowledgments}
	The author would like to thank J.M. Hoff da Silva for insightful discussions and for providing feedback on the manuscript.
	This work was supported by CNPq under the grants PQ-C 308651/2025-1 and 406997/2025-0.
	
	\section*{ORCID}
	
	\noindent Gustavo P. de Brito - \url{https://orcid.org/0000-0003-2240-528X}
	
	\appendix
	
	\section{Leading order flows around a free fixed-point \label{app:LO-Flow}}
	
	In Secs. \ref{subsec:ScalarELKO_2} and \ref{subsec:PauliInduced4Ferm}, we discussed the leading order behavior of renormalization group trajectories around a free fixed-point. Here, we provide further details on how this analysis was done.
	
	Typically, this kind of analysis relies on the linearized flow around a fixed point. In this case, the behavior of the flow in the neighborhood of a fixed point is dictated by the eigenvalues of the stability matrix, with components defined as
	\begin{equation}
		S_{ij}(\vec{g}_*) = \left( \frac{\partial \beta_{g_i}}{\partial g_j} \right)_{\vec{g}=\vec{g}_*} \,,
	\end{equation}
	where $\vec{g} = \{g_1,\cdots,g_n\}$ is a set of couplings, $\beta_{g_i}(\vec{g})$'s are the corresponding beta functions, and $\vec{g}_*$ denotes a fixed-point solution. For the systems investigated in Secs. \ref{subsec:ScalarELKO_2} and \ref{subsec:PauliInduced4Ferm} (in the approximation were we neglect the mass parameters), all beta functions are at least quadratic on the couplings, thus implying
	\begin{equation}
		S_{ij}(\vec{g}_* =\vec{0}) = 0 \qquad \text{$\forall$ $i,j$}.
	\end{equation}
	Therefore, the analysis based on linearized flows is not sufficient to give us information about the flow in the neighborhood of a fixed point. For this reason, we need to consider a different strategy to investigate the leading-order of flow around the free fixed-point.  
	
	To proceed with the leading-order analysis of the systems described in Secs. \ref{subsec:ScalarELKO_2} and \ref{subsec:PauliInduced4Ferm} around the free fixed-point, we use the $n$-dimensional \textit{blowing up method} ~\cite{Perko:2001}~. The basic idea is to express the couplings $\{g_i\}_{i=1,\cdots,n}$ in a spherical parametrization centered at the fixed point. For the free fixed-point, we can write 
	\begin{subequations}
		\begin{align}
			g_1(k) &= r(k) \cos \left(\phi_1(k)\right) \, ,\\
			g_2(k) &= r(k) \sin \left(\phi_1(k)\right) \, \cos \, \left(\phi_2(k)\right) \, ,\\
			&\qquad \vdots \\
			g_{n-1}(k) &= r(k) \sin \left(\phi_1(k)\right) \, \cdots \, \cos \left(\phi_{n-1}(k)\right) \, ,\\
			g_n(k) &= r(k) \sin \left(\phi_1(k)\right) \,\cdots \,\sin \left(\phi_{n-1}(k)\right) \, .
		\end{align}
	\end{subequations}
	where $r(k)$ is a scale-dependent radial coordinate and the $\phi_i(k)$'s denote scale-dependent angular coordinates. Using the new set of coordinates, we can rewrite the flow equations as
	\begin{subequations}
		\begin{align}
			k\partial_k r(k) &= \beta_{r}(r,\vec{\phi}) \, ,\\
			k\partial_k \phi_i(k) &= \beta_{\phi_i}(r,\vec{\phi}) \,,
		\end{align}
	\end{subequations}
	where $\beta_{r}$ and $\beta_{\phi_i}$`s denote the radial and angular beta functions, respectively. In the new set of coordinates, the free fixed-point correspond to $r_* = 0$.
	If the original set of beta functions are at least quadratic on the couplings, we can show that their ``spherical" counterparts take the form
	\begin{subequations}
		\begin{align}
			&\beta_{r}(r,\vec{\phi}) = r^2 \,H(\vec{\phi}) + \mathcal{O}(r^3) \, ,\\
			&\beta_{\phi_i}(r,\vec{\phi}) = r \, F_i(\vec{\phi}) + \mathcal{O}(r^2) \,,
		\end{align}
	\end{subequations}
	when expanded to leading order around the fixed point. To investigate non-trivial solutions connected with the free-fixed point, we search for invariant rays, denoted as $\vec{\phi}^\star = \{ \phi_{1}^\star , \cdots , \phi_{n-1}^\star\}$, and defined according to
	\begin{equation}
		F_i(\vec{\phi}^\star) = 0 \,,\quad\text{such that}\quad H(\vec{\phi}^\star)<0 \,.
	\end{equation}
	In principle, a set of renormalization group equations can support more than one invariant rays, each one defining a class of trajectories compatible with the free fixed-point.
	
	Expanding the renormalization group equations around an invariant ray and keeping only leading-order terms, we find  
	\begin{subequations}
		\begin{align}
			k\partial_k r(k) &= r(k)^2 \,H(\vec{\phi}^\star) \, , \label{eq:Radial_Flow_LO}\\
			k\partial_k \rho_i(k) &= \sum_{j=1}^{n-1} r(k) \,M_{ij}(\vec{\phi}^\star)\,\rho_j(k) \,, \label{eq:Angular_Flow_LO}
		\end{align}
	\end{subequations}
	where $M_{ij}(\vec{\phi}^\star)$ is defined as
	\begin{equation}
		M_{ij}(\vec{\phi}^\star) = \left(\frac{\partial \beta_{\phi_i}}{\partial \phi_j}\right)_{\vec{\phi} = \vec{\phi}^\star} \,.
	\end{equation}
	We can easily integrate Eqs. \eqref{eq:Radial_Flow_LO} and \eqref{eq:Angular_Flow_LO}, resulting in 
	\begin{subequations}
		\begin{equation}
			r(k) \approx \frac{C_0}{|H(\vec{\phi}^\star)|\,\log(k/k_0)} \, ,
		\end{equation}
		\begin{equation}
			\rho_i(k) \approx \sum_{l=1}^{n-1} \frac{C_l}{C_0} \, v_i^{(l)}(\vec{\phi}^\star) \big[\log(k/k_0)\,\big]^{\lambda_l(\vec{\phi}^\star)/|H(\vec{\phi}_0)|} 
		\end{equation}
	\end{subequations}
	where $\{C_0,\cdots,C_{n-1}\}$ is a set of integration constants, $v_i^{(l)}$ and $\lambda_l$ are defined by the eigenvalue equation
	\begin{eqnarray}
		\sum_{j=1}^{n-1} M_{ij}(\vec{\phi}^\star) v_j^{(l)}(\vec{\phi}^\star) = \lambda_l(\vec{\phi}^\star) v_i^{(l)}(\vec{\phi}^\star) \, ,
	\end{eqnarray}
	and $k_0$ denotes a reference scale. Note that $|H(\vec{\phi}^\star)| = - H(\vec{\phi}^\star)$, since $H(\vec{\phi}^\star) < 0$. Demanding $r(k) \geq 0$, we constrain $C_0 \geq 0$, where the equality implies a trivial flow.
	
	The invariant ray $\vec{\phi}^\star$ is UV-attractive from an \textit{eigendirections} $\vec{v}_l(\vec{\phi}^\star)$, if the corresponding eigenvalue $\lambda_l$ is negative. Let us suppose we label the eigenvalues such that $\{\lambda_1,\cdots,\lambda_p\}$ are all negative, and $\{\lambda_{p+1},\cdots,\lambda_{n-1}\}$ contains non-negative eigenvalues. Using this labeling, we can rewrite the solution for $\rho_i(k)$ as follows
	\begin{equation}
		\begin{aligned}
			\rho_i(k) &\approx 
			\sum_{l=1}^{p} \frac{C_l}{C_0} \, v_i^{(l)}(\vec{\phi}^\star) \big[\log(k/k_0)\,\big]^{-|\lambda_l(\vec{\phi}^\star)|/|H(\vec{\phi}_0)|}\\
			&+ \sum_{l=p+1}^{n-1} \frac{C_l}{C_0} \, v_i^{(l)}(\vec{\phi}^\star) \big[\log(k/k_0)\,\big]^{|\lambda_l(\vec{\phi}^\star)|/|H(\vec{\phi}_0)|}
		\end{aligned}
	\end{equation}
	If we demand the flow to approach the invariant ray $\vec{\phi}^\star$ in the UV limit $k \to \infty$, which translates into $\rho(k\to\infty) =0$, then we need to set $C_{p+1} = C_{p+2} = \cdots = C_{n-1} = 0$. The remaining integration constants are free parameters that need to be fixed experimentally.

	\bibliographystyle{ws-mpla}
	\bibliography{biblio}
	
\end{document}